\documentstyle[times,psfig,mathrsfs]{mn}

\def\4he{$^4$He}
\def\3he{$^3$He}
\def\7li{$^7$Li}

\def\yd{$y_{\rm D}$~}
\def\yo{$y_{\rm O}$~}

\def\hi{H\thinspace{$\scriptstyle{\rm I}$}}
\def\di{D\thinspace{$\scriptstyle{\rm I}$}}
\def\oi{O\thinspace{$\scriptstyle{\rm I}$}}

\def\be{\begin{equation}}
\def\ee{\end{equation}}

\newif\ifAMStwofonts
\AMStwofontstrue

\ifoldfss
  \ifCUPmtlplainloaded \else
    \NewTextAlphabet{textbfit} {cmbxti10} {}
    \NewTextAlphabet{textbfss} {cmssbx10} {}
    \NewMathAlphabet{mathbfit} {cmbxti10} {} 
    \NewMathAlphabet{mathbfss} {cmssbx10} {} 
  \fi
  \ifAMStwofonts
    \ifCUPmtlplainloaded \else
      \NewSymbolFont{upmath} {eurm10}
      \NewSymbolFont{AMSa} {msam10}
      \NewMathSymbol{\upi}     {0}{upmath}{19}
      \NewMathSymbol{\umu}     {0}{upmath}{16}
      \NewMathSymbol{\upartial}{0}{upmath}{40}
      \NewMathSymbol{\leqslant}{3}{AMSa}{36}
      \NewMathSymbol{\geqslant}{3}{AMSa}{3E}

      \let\leq=\leqslant \let\le=\leqslant
      \let\geq=\geqslant 
    \fi
  \fi
\fi 

\ifnfssone
  \newmathalphabet{\mathit}
  \addtoversion{normal}{\mathit}{cmr}{m}{it}
  \addtoversion{bold}{\mathit}{cmr}{bx}{it}
  \newmathalphabet{\mathbfit} 
  \addtoversion{normal}{\mathbfit}{cmr}{bx}{it}
  \addtoversion{bold}{\mathbfit}{cmr}{bx}{it}
  \newmathalphabet{\mathbfss} 
  \addtoversion{normal}{\mathbfss}{cmss}{bx}{n}
  \addtoversion{bold}{\mathbfss}{cmss}{bx}{n}
  \ifAMStwofonts
    \ifCUPmtlplainloaded \else
      \UseAMStwoboldmath
      \makeatletter
      \new@mathgroup\upmath@group
      \define@mathgroup\mv@normal\upmath@group{eur}{m}{n}
      \define@mathgroup\mv@bold\upmath@group{eur}{b}{n}
      \edef\UPM{\hexnumber\upmath@group}
      \new@mathgroup\amsa@group
      \define@mathgroup\mv@normal\amsa@group{msa}{m}{n}
      \define@mathgroup\mv@bold\amsa@group{msa}{m}{n}
      \edef\AMSa{\hexnumber\amsa@group}
      \makeatother
      \mathchardef\upi="0\UPM19
      \mathchardef\umu="0\UPM16
      \mathchardef\upartial="0\UPM40
      \mathchardef\leqslant="3\AMSa36
      \mathchardef\geqslant="3\AMSa3E

      \let\leq=\leqslant \let\le=\leqslant
      \let\geq=\geqslant 
    \fi
  \fi
\fi 

\ifnfsstwo
  \DeclareMathAlphabet{\mathbfit}{OT1}{cmr}{bx}{it}
  \SetMathAlphabet\mathbfit{bold}{OT1}{cmr}{bx}{it}
  \DeclareMathAlphabet{\mathbfss}{OT1}{cmss}{bx}{n}
  \SetMathAlphabet\mathbfss{bold}{OT1}{cmss}{bx}{n}
  \ifAMStwofonts
    \ifCUPmtlplainloaded \else
      \DeclareSymbolFont{UPM}{U}{eur}{m}{n}
      \SetSymbolFont{UPM}{bold}{U}{eur}{b}{n}
      \DeclareSymbolFont{AMSa}{U}{msa}{m}{n}
      \DeclareMathSymbol{\upi}{0}{UPM}{"19}
      \DeclareMathSymbol{\umu}{0}{UPM}{"16}
      \DeclareMathSymbol{\upartial}{0}{UPM}{"40}
      \DeclareMathSymbol{\leqslant}{3}{AMSa}{"36}
      \DeclareMathSymbol{\geqslant}{3}{AMSa}{"3E}

      \let\leq=\leqslant \let\le=\leqslant
      \let\geq=\geqslant 
    \fi
  \fi
\fi 

\ifCUPmtlplainloaded \else
  \ifAMStwofonts \else 
    \def\upi{\pi}
    \def\umu{\mu}
    \def\upartial{\partial}
  \fi
\fi

   \title[Connecting the primordial and Galactic deuterium
           abundances]{Connecting the primordial and Galactic deuterium
           abundances}
   \author[G. Steigman, D. Romano \& M. Tosi]{Gary Steigman$^{1}$, Donatella 
            Romano$^{2}$ and Monica Tosi$^{2}$\\
	    $^{1}$Departments of Physics and Astronomy, The Ohio State 
                  University, Columbus, OH, USA\\
            $^{2}$INAF\,--\,Osservatorio Astronomico di Bologna,
                  Via Ranzani 1, I-40127 Bologna, Italy}
   
\begin{document}

     \date{Accepted 2007 March 23. Received 2007 February 21; in original form 
           2006 December 18}

     \pagerange{\pageref{firstpage}--\pageref{lastpage}} \pubyear{2007}

     \maketitle

     \label{firstpage}


   \begin{abstract}
   The deuterium abundances inferred from observations of the interstellar 
   medium within 1~--~2 kpc of the Sun range over a factor of three and the 
   corresponding oxygen abundances show an even larger dispersion.  While the 
   lower D (and O) abundances likely result from depletion onto dust, the 
   higher D abundances are consistent with the BBN-predicted primordial D 
   abundance and chemical evolution models of the Galaxy with infall of 
   primordial or nearly primordial material.  The large ranges in deuterium 
   and oxygen abundances suggest that the effects of depletion and/or infall 
   have not been homogenized in the local interstellar medium.
   \end{abstract}



   \section{Introduction}
   \label{intro}

   Of the light nuclides produced in astrophysically interesting abundances 
   (D, \3he, \4he, \7li) during Big Bang Nucleosynthesis (BBN), deuterium 
   is the baryometer of choice for several reasons.  For example, the 
   BBN-predicted primordial deuterium abundance is sensitive to the baryon 
   (nucleon) density [$y_{\rm DP} \equiv 10^{5}$(D/H)$_{\rm P} \propto 
   \eta_{10}^{-1.6}$ where $\eta_{10} \equiv 
   10^{10}(n_{\rm B}/n_{\gamma})_{0}$] and the post-BBN evolution of D is 
   simple and monotonic: as gas cycles through stars, deuterium is only 
   destroyed (Epstein, Lattimer \& Schramm 1976; Prodanovi\'c \& Fields 2003). 
   A consequence of this simple evolution is that the post-BBN deuterium 
   abundance is a probe of the fraction of the interstellar gas which has 
   never been cycled through stars (Steigman \& Tosi 1992, 1995).  Models of 
   the chemical evolution of the Galaxy, constrained by observations of the 
   stellar and gas phase metal abundances, follow the history of the gas.  As 
   a bonus they predict the time-evolution of the deuterium abundance.  Until 
   relatively recently observations of deuterium in the Universe were limited 
   to the interstellar medium (ISM) of the Galaxy, so that determinations of 
   primordial deuterium required the input of chemical evolution models and, 
   hence, were model-dependent (Steigman \& Tosi 1992, 1995; Vangioni-Flam, 
   Olive \& Prantzos 1994; Prantzos 1996; Tosi 1996; Scully et al. 1997; Tosi 
   et al. 1998).  This changed with the advent of large, ground-based 
   telescopes and with the Hubble Space Telescope (HST), which have enabled 
   determinations of the D abundance along the lines-of-sight (LOS) to a few 
   (at present, six) high-redshift, low-metallicity QSO Absorption Line 
   Systems (QSOALS; Kirkman et al. 2003; O'Meara et al. 2006).  At the same 
   time observations of the angular distribution of the temperature 
   fluctuations in the Cosmic Background Radiation (CBR), in combination with 
   new data on the large-scale structure of the Universe (see, e.g., Spergel 
   et al. 2006; Tegmark et al. 2006), have constrained the baryon abundance to 
   $\sim 3~\%$, $\eta_{10} = 6.08 \pm 0.19$, leading to a BBN-predicted 
   primordial D abundance accurate to $\sim$~5 -- 6~\%, $y_{\rm DP} = 
   2.61\pm0.15$.  Starting with this primordial abundance, what do Galactic 
   chemical evolution (GCE) models predict for the current local ISM deuterium 
   abundance and how well do these predictions compare to observations?

   This paper is organized as follows. In Section~2 the predicted and observed 
   abundances of D and O in the local ISM are reviewed and compared, 
   accounting for depletion onto dust grains as well as localized infall of 
   metal-poor, D-enhanced gas.  The implications of ISM inhomogeneities are 
   further discussed in Section~3.  Finally, in Section~4 we draw our 
   conclusions.

   \section{Comparing ISM Predicted And Observed Abundances}

   As gas cycles through stars, D is destroyed and the abundances of the
   ``metals", in particular oxygen, increase (from zero primordially).  In 
   addition to star formation, gas flows play a fundamental r\^ole in 
   determining the chemical composition of gas and stars in galaxies.  In 
   particular, infall of primordial, or nearly primordial, gas plays a crucial 
   r\^ole in the chemical evolution of our own Galaxy, enhancing the ISM D 
   abundance, restoring it closer to its primordial value (see, e.g., Tosi 
   1988a, b).  As a result the ISM deuterium abundance is not expected to 
   depart from its primordial value until the ISM metallicity increases to 
   $\ga 0.1$ solar (Romano et al. 2006).  Underlying GCE models is the 
   ``conventional wisdom" that the local ISM is chemically homogeneous (i.e. 
   the timescale for mixing in the local ISM is much shorter than 
   stellar/chemical evolution timescales).  Consequently, GCE models predict 
   mean abundances for deuterium and the metals, e.g. oxygen, in the 
   vicinity of the Sun at the present time.  The data obtained from the Far
   Ultraviolet Satellite Explorer (FUSE; Oliveira et al. 2006; Oliveira 
   \& H\'ebrard 2006), however, reveal a factor three range for the D 
   abundance and an even larger range for the O abundance, clearly in conflict 
   with this simplifying assumption.  The gas phase abundances of D and O are 
   currently not distributed homogeneously in the local ISM.

   \subsection{Predicted ISM abundances}

   The GCE model-predicted ISM abundance of D is related to the primordial D 
   abundance through the ``astration" factor, $f_{\rm D} \equiv y_{\rm 
   DP}/y_{\rm D}^{\rm ISM}$.  Many GCE models are found in the literature 
   which address D evolution in the Milky Way.  Generally, they adopt 
   different prescriptions for the star formation rate, infall and outflow 
   rates, initial mass function (IMF) and stellar lifetimes (see, e.g., 
   Vangioni-Flam et al. 1994; Prantzos 1996; Tosi 1996; Scully et al. 1997; 
   Tosi et al. 1998).  Successful models which account for a wealth of 
   observational data always find ``low'' values for $f_{\rm D}$, in the range 
   $\sim 1.5-3$.  Recent models (Romano et al. 2006) in which the 
   prescriptions of the stellar IMF and stellar lifetimes (Scalo 1986; Kroupa, 
   Tout \& Gilmore 1993; Maeder \& Meynet 1989; Schaller et al. 1992) are 
   allowed to change, predict that the astration factor and the current ISM 
   oxygen abundances lie in the ranges, $1.39~\la f_{\rm D}~\la 1.83$ and 
   $7~\la y_{\rm O}~\la 12$, where $y_{\rm O} \equiv 10^{4}$(O/H).  For a 
   BBN-predicted primordial D abundance of $y_{\rm DP} = 2.61$, the 
   GCE-predicted ISM D abundance is expected to lie in the range, $1.43~\la 
   y_{\rm D}^{\rm ISM}~\la 1.88$.  

   There are at least two physical effects capable of accounting for the 
   observed variations in D and O abundances along different LOS within $\sim 
   1-2$~kpc of the Sun.  For an explanation of the variable deuterium 
   abundances Linsky et al. (2006) embrace the suggestion of Draine (2004), 
   based on earlier work of Jura (1982), that deuterium may be preferentially 
   depleted (compared to hydrogen) onto dust grains.  While Linsky et al. do 
   not address the variation in oxygen abundances, depletion might provide a 
   viable explanation for the dispersion among the observed oxygen abundances 
   as well.

   Another possibility is recent infall to the disk of the Galaxy of 
   unprocessed (D-rich, O-poor) material which has been incompletely mixed in 
   the ISM (see also Geiss, Gloeckler \& Charbonnel 2002).  In this case there 
   would be an anti-correlation between the D and O abundances observed along 
   such contaminated LOS.  For example, if $y_{\rm O}^{\rm ISM}$ and 
   $y_{\rm D}^{\rm ISM} \equiv y_{\rm DP}/f_{\rm D}$ are the ``true" ISM 
   abundances and $y_{\rm O}^{\rm OBS}$ and $y_{\rm D}^{\rm OBS}$ the 
   abundances observed along a specific LOS, then incompletely mixed infall 
   predicts
   \be
   y_{\rm D}^{\rm OBS} = y_{\rm DP} + (y_{\rm D}^{\rm ISM} - 
   y_{\rm DP})y_{\rm O}^{\rm OBS}/y_{\rm O}^{\rm ISM}.
   \ee
   For example, if the Scalo (1986) IMF along with the stellar lifetimes of 
   Schaller et al. (1992) are adopted, then according to Romano et al. (2006), 
   $f_{\rm D} = 1.39$ and $y_{\rm O}^{\rm ISM} = 8.4$, so that 
   \be
   y_{\rm D}^{\rm OBS} = 2.61 - 0.087y_{\rm O}^{\rm OBS}
   \label{dvsoinf}
   \ee
   is predicted.  

   Since FUSE is often unable to provide sufficiently accurate \hi~column 
   densities, to interpret these data people usually rely on \hi~column 
   density determinations from other telescopes/observers.  As a result, it is 
   possible that some of the dispersion among the observed D and O abundances 
   may be due to systematic errors in the associated \hi~column densities 
   (Timmes et al. 1997; H\'ebrard \& Moos 2003; Steigman 2003).  However, 
   the ratio of D to O is unaffected by such errors.  For 10(D/O) = 
   $y_{\rm D}/y_{\rm O}$, eq.~\ref{dvsoinf} predicts
   \be
   10({\rm D}/{\rm O}) = 26.1/y_{\rm O}^{\rm OBS} - 0.87.
   \label{dovsoinf}
   \ee

   Of course, it may well be the case that both mechanisms (depletion and 
   infall) are at work in the ISM.  If so, the anticorrelations described by 
   equations $1-3$ provide an {\it upper} envelope to the observations in the 
   \yd $-$ \yo and D/O -- \yo planes (see Figs.~\ref{fig:deplet} and 
   \ref{fig:infall}).  Infall of (nearly) primordial gas is often invoked to 
   account for abundance variations in the ISM (e.g. Cartledge et al. 2004; 
   Knauth, Meyer \& Lauroesch 2006, and references therein).  The required 
   infall episodes should have occurred recently enough to avoid complete 
   mixing of the ISM.  The exact timescale on which clouds of partially 
   ionized gas falling into the magnetized ISM will be thoroughly homogenized 
   is not clear.  At the same time it is clear that if depletion is the 
   correct explanation for the observed variations in D and O abundances, then 
   local depletions have not been homogenized either. Therefore, although it 
   will be seen below that recent inhomogeneous infall on small (pc-sized) 
   scales is {\it not required} to achieve concordance between the observed 
   ISM abundances and GCE models, neither should it be rejected out of hand as 
   contributing to the observed dispersion among the gas phase D and O 
   abundances.

   \subsection{Observed -- gas phase -- ISM abundances}

   The abundance of deuterium in the relatively local ISM (within $1-2$~kpc of 
   the Sun) has been explored for some 30 years beginning with observations by 
   the~{\it Copernicus} satellite and continuing with a series of space-based 
   UV detectors (e.g. {\it IUE}, {\it GHRS} and~{\it STIS} on the HST) up to 
   the current~{\it IMAPS} instrument on the FUSE satellite.  The most recent 
   and comprehensive abundance analyses are based on the data acquired by  
   FUSE (Oliveira et al. 2006; Linsky et al. 2006; Oliveira \& H\'ebrard 
   2006).  These data are adopted in our analysis here, together with the 
   abundances in the warm neutral medium of the lower Galactic halo recently 
   determined by Savage et al. (2007). 

\begin{figure}
   \psfig{figure=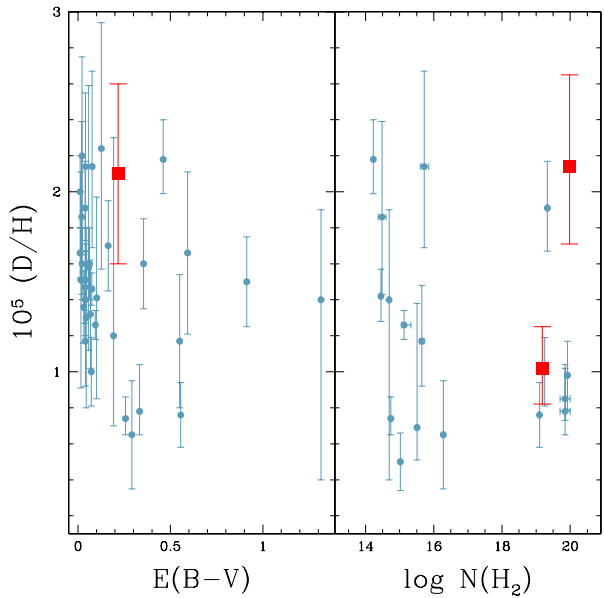,width=\columnwidth}
      \caption{D/H versus E(B$-$V) (left panel) and logN(H$_2$) (right panel). 
      Estimates of E(B$-$V) are from Schlegel et al. (1998); H$_{2}$ data 
      are from Hoopes et al. (2003), Linsky et al. (2006) and Oliveira \& 
      H\'ebrard (2006); D/H determinations are taken from Oliveira et al. 
      (2006; dots) and Oliveira \& H\'ebrard (2006; squares).}
      \label{fig:ext}
   \end{figure}

   Linsky et al. (2006) suggest that local variations in the observed 
   gas-phase D/H ratios can be explained by spatial variations in the 
   depletion of deuterium onto dust grains.  While this hypothesis of 
   localized depletion provides a reasonable explanation for the observed 
   variations in the D and O abundances, there is little direct evidence in 
   support of it.  For example, although the FUSE observations cover a range 
   of four orders of magnitude in \hi~column density [$17.6~\la$ 
   logN(\hi)$~\la$ 21.2], both the lowest and the highest deuterium 
   abundances are found for 19.2$~\la$ logN(\hi)$~\la$ 21.0 (see Oliveira 
   et al. 2006, their fig.~17, upper panel, and Oliveira \& H\'ebrard 2006).  
   Of course, since the LOS explored by FUSE range in distance (to the 
   background star) from tens of parsecs to $1-2$~kpc, high \hi~column density 
   is not equivalent to high gas density.  With few exceptions the FUSE data 
   do not permit a direct determination of the local gas density.  A surrogate 
   for the gas density, $n$(\hi)$ \equiv $N(\hi)/$d$, where $d$ is the 
   distance to the background illuminating star, ranges over nearly three 
   orders of magnitude, from $\sim 0.002$~cm$^{-3}$ to $\sim 0.6$~cm$^{-3}$.  
   While this estimate doesn't preclude much higher local densities, the 
   largest of these is smaller than the densities associated with molecular 
   clouds where depletion is expected to be efficient.  Furthermore, here too, 
   the evidence in support of D-depletion is ambiguous in that although some 
   of the lowest (depleted?) D abundances are found at the highest values of 
   $n$(\hi), so too are the majority of the highest (undepleted?) D abundances 
   (see Oliveira et al. 2006, their fig.~19, upper panel).  Alternatively, if 
   dust depletion is, indeed, the culprit it might be anticipated that the 
   deuterium and oxygen abundances should be {\it anticorrelated} with the LOS 
   reddening, as measured by E(B$-$V) or, with the molecular hydrogen 
   fraction.  We explored these possibilities (see Fig.~\ref{fig:ext} for 
   deuterium) using estimates of E(B$-$V) from Schlegel, Finkbeiner \& Davis 
   (1998) and H$_{2}$ data from Hoopes et al. (2003), Linsky et al. (2006), 
   Oliveira \& H\'ebrard (2006). We find that although there is a tendency for 
   the lowest (highest) deuterium abundances to correspond to the highest 
   (lowest) E(B$-$V) values and H$_{2}$ fractions, many of the least reddened 
   LOS, as well as those with the smallest H$_{2}$ fractions, where depletion 
   may be less likely, also have {\it low} D abundances.  Furthermore, two out 
   of eight LOS with the highest H$_{2}$ fractions (HD\,41161 and 
   PG\,0038+199) have some of the highest D abundances observed.  Similar 
   trends are found for the oxygen abundances [low O/H at low E(B$-$V), low 
   H$_{2}$ fraction].
   

\begin{figure}
   \psfig{figure=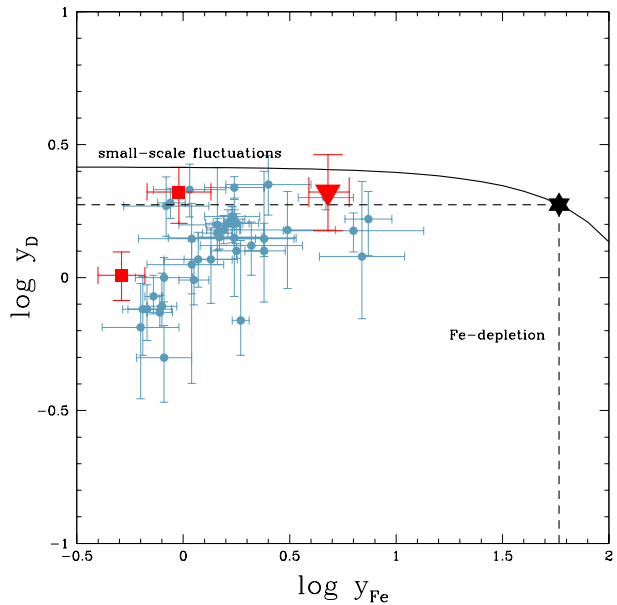,width=\columnwidth}
      \caption{ISM deuterium abundances versus iron abundances $y_{\rm Fe} 
      \equiv 10^{6}$(Fe/H).  Data from: Linsky et al. (2006) for Fe and 
      Oliveira et al. (2006) for D (dots); Oliveira \& H\'ebrard (2006; full 
      squares); Savage et al. (2007; upside-down triangle).  The star shows 
      the abundances predicted by the fiducial chemical evolution model (see 
      the text).  D and Fe depletion require that the data lie below and to 
      the left of the dashed lines.  The solid curve shows the effect of 
      local abundance fluctuations resulting from incompletely mixed infall 
      of primordial (D-enhanced, Fe-free) material.}
      \label{fig:dvsfe}
   \end{figure}

   Perhaps the strongest evidence in support of D depletion comes from the 
   correlation between D and Ti abundances (Prochaska, Tripp \& Houk 2005; but 
   see also H\'ebrard et al. 2005).  The FUSE data (Oliveira et al. 2006; 
   Linsky et al. 2006; Oliveira \& H\'ebrard 2006) reveal a similar trend 
   between D and Fe.  These data, plotted in Fig.~\ref{fig:dvsfe}, show that 
   the lowest D abundances are, indeed, correlated with the lowest Fe 
   abundances [log~$y_{\rm Fe} ~\la 0.0$, where $y_{\rm Fe} \equiv 
   10^{6}$(Fe/H)].  However, these data also suggest that deuterium may be 
   undepleted along several of the LOS for which log~$y_{\rm Fe}~\ga 0.0 - 
   0.2$.  This calls into question the Linsky et al. (2006) {\it assumption} 
   that a lower limit to the ``true" ISM D abundance is provided by the 
   \emph{maximum} of $y_{\rm D}^{\rm OBS}$.
   

   \begin{figure}
   \psfig{figure=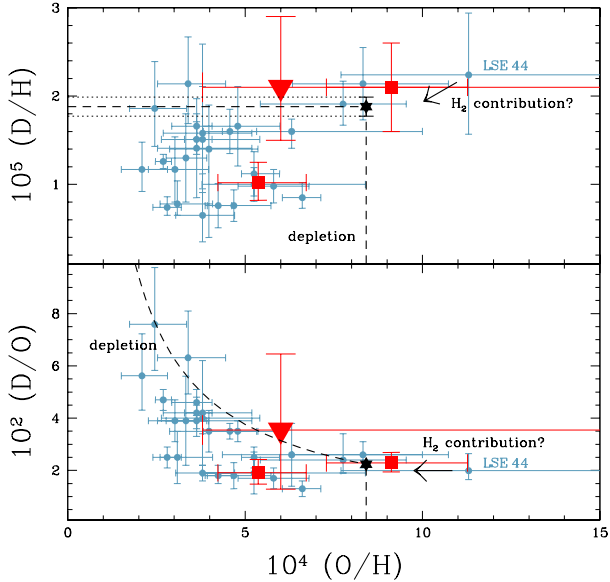,width=\columnwidth}
      \caption{ The upper panel shows the deuterium versus oxygen abundances.  
      The lower panel plots the ratio of deuterium to oxygen as a function of 
      the oxygen abundances.  In both panels, the FUSE data are displayed as 
      filled circles (Oliveira et al. 2006), squares (Oliveira \& H\'ebrard 
      2006) and the upside-down triangle (Savage et al. 2007), while the star 
      indicates the ISM abundances predicted by the fiducial chemical 
      evolution model (see the text).  The horizontal dashed and dotted lines 
      in the upper panel are the predicted ISM deuterium abundance and its 
      $\pm 1\sigma$ uncertainty, respectively. The vertical dashed lines in 
      both panels correspond to the predicted ISM oxygen abundance. The dashed 
      curve in the lower panel shows the maximum values of D/O if O is 
      depleted, but D is not.  Data below and to the left of the dashed lines 
      are consistent with D and O depletion.  The arrows in each panel 
      indicate the effect of a possible contribution of molecular hydrogen to 
      the D and O abundances for the LOS towards LSE\,44.}
      \label{fig:deplet}
   \end{figure}

\begin{figure}
   \psfig{figure=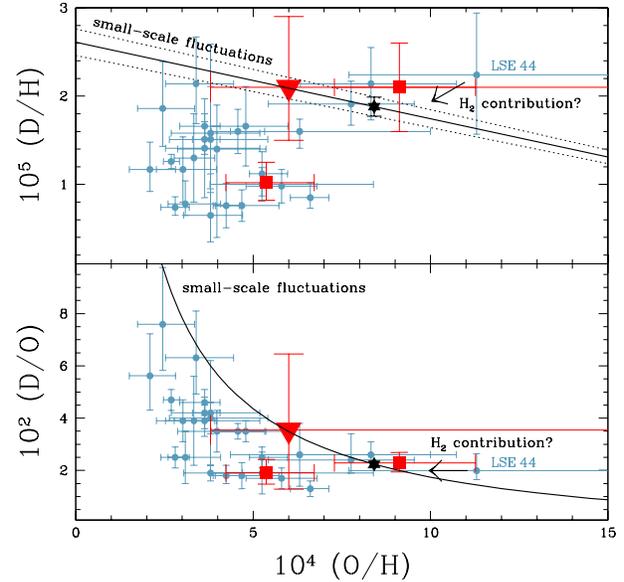,width=\columnwidth}
      \caption{ Deuterium abundances and D/O ratios as in 
      Fig.~\ref{fig:deplet}.  The solid curves show the effect of local 
      fluctuations in the abundances in the absence of depletion but 
      accounting for infall of primordial (D-enhanced, O-free) gas on small 
      spatial scales.}
      \label{fig:infall}
   \end{figure}

   If dust depletion is the dominant mechanism responsible for the observed 
   dispersion among the gas-phase abundances of deuterium and oxygen, the D 
   and O abundances should be {\it correlated} (as D appears to be with Ti and 
   Fe).  Incompletely mixed infall of unprocessed material would give rise to 
   an {\it anticorrelation} between the D and O abundances.  To explore these 
   trends in the FUSE data we concentrate on those LOS with data (mostly from 
   Oliveira et al. 2006) for both the D and O abundances.  The D/H and D/O 
   ratios versus O/H are plotted in Figs.~\ref{fig:deplet} and 
   \ref{fig:infall}.  If depletion dominates, none of the observed abundances 
   should exceed the GCE-predicted values.  For the fiducial model we adopted 
   the Scalo (1986) IMF along with the Schaller et al. (1992) stellar 
   lifetimes, so that $f_{\rm D} = 1.39$ and $y_{\rm O}^{\rm ISM} = 8.4$, 
   leading to $y_{\rm D}^{\rm ISM} = y_{\rm DP}/f_{\rm D} = 1.88\pm0.11$.  The 
   upper panel of Fig.~\ref{fig:deplet} shows D/H versus O/H along with the 
   upper bounds to their GCE-predicted ISM abundances if depletion dominates.  
   We note that while most of the LOS have D or O abundances which lie below 
   these bounds, consistent with depletion, none of them are more than 
   1$\sigma$ above these GCE-predicted ISM abundances, consistent with some of 
   these LOS being depletion-free.  One possible exception may be the LOS 
   towards LSE\,44 (Friedman et al. 2006; see Fig.~\ref{fig:deplet}).  
   However, as Liszt (2006) notes, it is important to account for molecular 
   hydrogen when determining the ISM deuterium abundance, since failure to do 
   so biases the resulting D/H (and O/H) ratio(s) upward.  Liszt notes that 
   molecular hydrogen is unlikely to be negligible (compared to \hi) whenever 
   N(\hi)$~> 4\times 10^{20}$cm$^{-2}$ and should be considered whenever 
   N(\hi)$~> 1.4\times 10^{20}$cm$^{-2}$.  This makes the LOS to LSE\,44, with 
   N(\hi)$ = 3.3\times 10^{20}$cm$^{-2}$, a good candidate for possibly 
   overestimated D and O abundances\footnote{ Although within the errors the D 
   and O abundances measured towards LSE\,44 are, in fact, consistent with the 
   predictions of our fiducial GCE model ($y_{\rm D}^{\rm ISM} = 1.88$, 
   $y_{\rm O}^{\rm ISM} = 8.4$), it is worth emphasizing that if molecular 
   hydrogen were to contribute N(H$_2$)/N(\hi)~= 0.19 along this LOS, then 
   $y_{\rm D}$ would be reduced from 2.24 to 1.88, and $y_{\rm O}$ from 11.3 
   to 9.5, in even better agreement with the model predictions.}.  This 
   possibility is indicated by the arrows in Figs.~\ref{fig:deplet} and 
   \ref{fig:infall}; note that the D/O ratios are unaffected.

   The large difference in \hi~and \di~column densities, $\sim 5$ orders of 
   magnitude, is a potential source of systematic errors for the D/H 
   determinations.  The difficulties associated with determining D/H may be 
   avoided by measuring the D/O ratio.  Since charge exchange reactions 
   between singly-ionized and neutral H and O (Field \& Steigman 1971) couple 
   \hi~to \oi, oxygen (\oi) is a good tracer of hydrogen (\hi), so that the 
   D/O ratio might be a good proxy for the D/H ratio (Timmes et al. 1997; 
   H\'ebrard \& Moos 2003; Steigman 2003).  The D/O ratio is plotted (versus 
   O/H) in the lower panel of Fig.~\ref{fig:deplet}.  Indeed, the FUSE data 
   (Oliveira et al. 2006; Linsky et al. 2006; Oliveira \& H\'ebrard 
   2006) reveal that the D/O ratio has an even larger dispersion than that 
   for D/H (a factor of $\sim$ 5, similar to that for O/H), suggesting that 
   the variations in O and in D abundances may be real and not the result of 
   systematic errors.  However, note that, as seen in the lower panel of 
   Fig.~\ref{fig:deplet}, the upper envelope of the D/O versus O/H data is 
   consistent with an undepleted ISM D abundance of $y_{\rm D}^{\rm ISM} = 
   1.88$.


   While the lower D and O abundances shown in Figs.~\ref{fig:deplet} and 
   \ref{fig:infall} may have resulted from the depletion of gas phase D and O, 
   the highest deuterium and oxygen abundances derived from the FUSE data are 
   consistent with present epoch average ISM deuterium and oxygen abundances 
   of $y_{\rm D}^{\rm ISM} = 1.88\pm0.11$ and $y_{\rm O}^{\rm ISM} = 8.4$.  
   For deuterium, these are consistent with a primordial abundance $y_{\rm DP} 
   = 2.61\pm0.15$ and a Galactic astration factor of 1.39, suggesting that gas 
   phase D (and O) may be undepleted along at least some LOS in the local 
   ISM.  Modest depletion of deuterium along some of these high-D LOS could 
   also be consistent with the data provided that the D (and O) variations are 
   due, in part, to the contribution of incompletely mixed, D-enhanced, 
   O-reduced infall\footnote{ Notice that our models are computed assuming 
   infall of primordial (i.e. O-free) material.  The conclusions are unchanged 
   if the infalling material is slightly enriched ($Z_{\mathrm{infall}} \le$ 
   0.2 $Z_\odot$, as for instance that coming from the Magellanic Stream; Tosi 
   1988b; Tosi et al. 1998).}. This latter possibility is shown in the two 
   panels of Fig.~\ref{fig:infall}.  Even including the LOS to LSE\,44, whose 
   D/H and O/H ratios might be overestimated if there is a non-negligible 
   contribution from unobserved H$_{2}$, the FUSE data are consistent with 
   primordial deuterium, GCE models, depletion, and -- possibly, but not 
   necessarily -- local fluctuations due to infall.

   \section{Discussion}

   Despite the evidence of 30 years of data to the contrary, until recently 
   the conventional wisdom has been that the local ISM is well mixed, with 
   unique abundances of the chemical elements, D and O in particular.  The 
   FUSE data, complementing earlier observations, which reveals a factor of 
   three range for D/H ($0.7~\la y_{\rm D}~\la 2.2$) and an even larger 
   dispersion among the O/H ratios ($2~\la y_{\rm O}~\la 8-11$), lays to rest 
   this overly simplified assumption.  While Linsky et al. (2006) don't 
   address the spread in oxygen abundances, they invoke localized depletion to 
   account for the observed range of deuterium abundances.  Although 
   independent observational evidence for depletion is currently ambiguous, 
   this mechanism does provide a description of the data.  If so, the question 
   becomes which, if any, of the observed D/H ratios represents the ``true" 
   (undepleted) ISM abundance?  

   Linsky et al. (2006) assume that the true D abundance is ``equal to or 
   slightly above the highest measured (D/H)$_{\rm gas}$ ratios".  Limiting 
   themselves to the 5 LOS with the highest D abundances, Linsky et al. 
   conclude that $y_{\rm D}^{\rm ISM} \geq 2.17\pm0.17$ (or, including a 
   correction for the Local Bubble D abundance, $y_{\rm D}^{\rm ISM} \geq 
   2.37\pm0.24$ or $y_{\rm D}^{\rm ISM} = 2.31\pm0.24$, depending on the 
   scheme for weighting the data).  These choices correspond to an astration 
   factor $\sim 1.1-1.2$, considerably smaller than those predicted by GCE 
   models (Tosi 1996 and references therein; Chiappini, Renda \& Matteucci
   2002; Romano et al. 2006).  However, given the errors associated with the 
   FUSE D/H determinations, it is unclear why their estimate should have been 
   limited to the 5 LOS with the highest central values of the D/H ratios 
   since there are more than a dozen other LOS whose central values of D/H 
   are within 1\,$\sigma$ of them.  To test our hypothesis that an astration 
   factor of 1.39, in combination with a BBN expected primordial abundance 
   of $y_{\rm DP} = 2.61\pm0.11$, is consistent with the FUSE data, we have 
   considered the 18 FUSE LOS with the highest D/H ratios ($y_{\rm D} \geq 
   1.5$).  Of these, nine have central values $y_{\rm D} \geq 1.9$ and nine 
   have $y_{\rm D} \leq 1.7$.  Using these data we may test the assumption 
   that the undepleted ISM D abundance is consistent with $y_{\rm D}^{\rm 
   ISM} = 1.88$.  We find that 13/18 (72\%) have central values of D/H 
   within 1\,$\sigma$ of our adopted ISM abundance and none are more than 
   1.4\,$\sigma$ away.  The reduced $\chi^{2}$ is 0.70 (for 17 degrees of 
   freedom) suggesting a more than $\sim 70$\% probability that the D 
   abundances along these 18 LOS are consistent with our adopted value 
   ($y_{\rm D}^{\rm ISM}$~= 1.88).  
   
   Indeed, we have been ``conservative" here in that, had we adopted for the 
   primordial D abundance the O'Meara et al. (2006) estimate of $y_{\rm DP} = 
   2.84$ inferred from the data for 6 high-z, low-Z QSOALS, consistency with 
   the FUSE data is improved, allowing a slightly larger astration factor of 
   $f_{\rm D} \sim 1.5$.
   
   In addition to inhomogeneously distributed depletion and possible 
   systematic errors in some of the \hi~column density determinations, some of 
   the variations in D and O abundances could be due to incompletely mixed, 
   recent infall of nearly primordial gas.  As shown in the lower panel of 
   Fig.~\ref{fig:infall}, the upper envelope of the D/O ratios reveals a trend 
   with O/H consistent with localized infall.  However, although infall 
   \emph{may} contribute to the large dispersions in the D and O abundances, 
   depletion appears to be \emph{required} to account for the lowest D/H and 
   O/H observed ratios.  In any case, with or without localized infall, at 
   least some of the current GCE models are consistent with the FUSE-inferred 
   and CBR-predicted BBN D abundances.

   \section{Summary and Conclusions}

   Standard galactic chemical evolution models predict the climate and not the
   weather and, hence, they provide \emph{average} abundances for deuterium 
   and oxygen (as well as other chemical species) in the local ISM. This is in 
   contrast to the large dispersion among the gas phase D and O abundances in 
   the neutral ISM revealed by 30 years of data.  Here we have demonstrated, 
   despite recent claims to the contrary (Linsky et al. 2006; Savage et al. 
   2007), that current models for the chemical evolution of the Milky Way 
   (Romano et al. 2006) are consistent with the FUSE inferred and 
   CBR-predicted BBN deuterium abundances.  While depletion of D (and O) onto 
   dust grains (Draine 2004) alone may suffice to account for the observed 
   variations, recent infall of (nearly) primordial gas on small spatial 
   scales can not be ruled out as a viable explanation for at least some of 
   the highest observed D abundances.  It is worth emphasizing that such 
   inhomogeneous infall -- which must have occurred recently enough that 
   mixing processes have not yet erased its elemental abundance signature -- 
   sums up to the continuous, homogeneous infall required by chemical 
   evolution models to account for the solar neighbourhood metallicity data.  
   Finally, notice that neglect of possible contributions from molecular 
   hydrogen (Liszt 2006) may have biased some of the observed abundances 
   upwards, enlarging the observationally inferred range for the present-day 
   D/H (and O/H).

   \section*{Acknowledgments}
   This research is supported at The Ohio State University by a grant 
   (DE-FG02-91ER40690) from the US Department of Energy. DR and MT 
   acknowledge useful discussions with the members of the LoLa-GE Team, 
   the warm hospitality and generous financial support at the International 
   Space Science Institute in Bern.

\bsp

\label{lastpage}

\end{document}